\documentclass[10pt]{article}
\usepackage{graphicx}
\usepackage{amssymb,amsmath}

\begin{document}

\title{Correction to ``A Note on Gallager's Capacity Theorem for Waveform Channels"}

\author{Edwin Hammerich\\
    \small Ministry of Defence\\
    \small Kulmbacher Str. 58--60, D-95030 Hof, Germany\\
    \small Email: edwin.hammerich@ieee.org}
\date{}
\maketitle

\begin{abstract}
We correct an alleged contradiction to Gallager's capacity theorem for waveform channels as presented in 
a poster at the 2012 IEEE International Symposium on Information Theory.
\end{abstract}

\newtheorem{definition}{Definition}
\newtheorem{theorem}{Theorem}
\newtheorem{proposition}{Proposition}
\newtheorem{lemma}{Lemma}
\newtheorem{remark}{Remark}
\newcommand{\e}{\mathrm{e}}

\section{Brief Description of the Poster \cite{Poster}}\label{Section_I}
Gallager's capacity theorem \cite[Theorem 8.5.1]{Gallager} is literally reproduced on p.~1 of the 
poster along with the figure \cite[Figure 8.5.1]{Gallager} plus some explanations taken 
from Gallager's book \cite{Gallager}. On p.~2, the so-called Gallager channel (now called Gaussian 
waveform channel), i.e., a Gaussian filter with additive white Gaussian noise (AWGN), is treated as a 
special instance of Gallager's theorem. On p.~3, the heat channel, a linear time-varying (LTV) filter 
with AWGN, is depicted followed by a characterization of its capacity by water-filling 
in the time-frequency plane \cite[Theorem 2]{Hammerich2}. Finally, on p.~4, an example of a signal 
transmission is given aiming at underpinning the apparent contradiction to Gallager's theorem in Figs.~3 
and 4 of the poster. The curves for 
the heat channel in the latter two figures are incorrect; the correct curves are provided in the present 
publication.\footnote{Still, it remains to be remarked that Theorem 8.5.1 in \cite{Gallager} is 
improperly stated in so far as the time parameter $T$ actually tends to infinity (see the proof of the 
theorem in \cite{Gallager})} 

\section{Capacity of the Heat Channel Revisited}
Evaluation of the double integrals occurring in the water-filling theorem \cite[Theorem 2]{Hammerich2} 
(cf. p. 3 of \cite{Poster}) and subsequent elimination of the parameter $\nu$ results in the 
closed-form representation of the capacity (in nats per transmission) of the heat channel, namely
\begin{equation}
  C(S)=\frac{\alpha\beta}{2}\left[w_0\left(\frac{S}{(\alpha\beta/2)\theta^2}
                \right)\right]^2+o(\alpha\beta)\,(\alpha\beta\rightarrow\infty),  \label{Eq1}
\end{equation}
where $y=w_0(x)$ is the inverse function of $y=(2x-1)\e^{2x}+1,\,x\ge 0$, and $o(\cdot)$ is the standard 
Landau little-o symbol. Eq.~(\ref{Eq1}) coincides with the capacity formula \cite[Eq. (9)]{Hammerich1} 
(as it should).

\section{Resolving the Contradiction}

\begin{figure}
  \centering
  \includegraphics[width=2.5in]{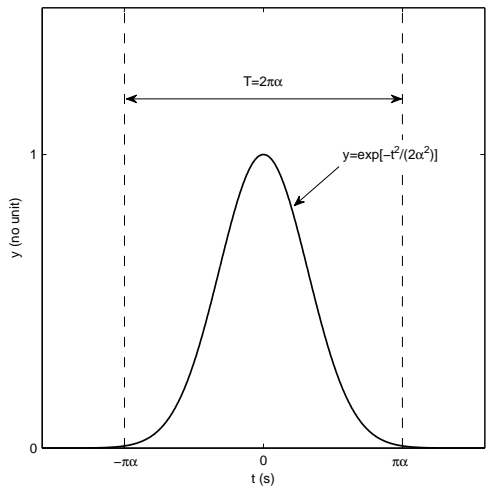}
  \caption{Time window at the input and output side of the heat channel's LTV filter.}
  \label{Figure_1}
\end{figure}

By reason of the Gaussian time window (see Fig.~\ref{Figure_1}) at the input and output side of the LTV 
filter (see \cite[Fig. 5]{Poster} and \cite[Eq. (4)]{Hammerich2}, resp.), capacity achieving 
input/output signals for the heat channel will have the approximate duration $T=2\pi\alpha$ (at least 
when the input energy is sufficiently high [but not too high!]). Therefore, transition to capacity as a 
rate should be done as follows. Put for the average input energy $S=2\pi\alpha P$, where $P$ is average 
power. Then, form the time average $\lim_{\alpha\rightarrow\infty}C(S)/(2\pi\alpha)$. Because of 
Eq.~(\ref{Eq1}) it asssumes the value
\begin{equation}
   \bar{C}(\mathrm{SNR})=\frac{1}{2\pi}\cdot W[w_0(2\pi\cdot2\mathrm{SNR})]^2\log_2\e
                                             \;\;\mbox{(bit/s)},              \label{Eq2}
\end{equation}
where $\mathrm{SNR}=P/(WN_0)$, $W=\beta/2$ is the approximate bandwidth in positive frequencies 
measured in hertz (cf. \cite[Fig. 2]{Poster}), and $N_0=2\theta^2$ is the one-sided noise power spectral 
density of the AWGN. Similarly, for $\bar{C}(\mathrm{SNR})/W$ as a function of $E_\mathrm{b}/N_0$ we 
obtain the parametric representation
\begin{eqnarray}
   \frac{E_\mathrm{b}}{N_0}&=&\frac{\mathrm{SNR}}{\frac{1}{2\pi}[w_0(4\pi\mathrm{SNR})]^2
                                                                           \log_2\e}\label{Eq3a}\\
   \frac{\bar{C}(\mathrm{SNR})}{W}&=&\frac{1}{2\pi}[w_0(4\pi\mathrm{SNR})]^2
                                                           \log_2\e\;\;(\mbox{bit/s/Hz})\label{Eq3b}.
\end{eqnarray}

\begin{figure}
  \centering
  \includegraphics[width=2.5in]{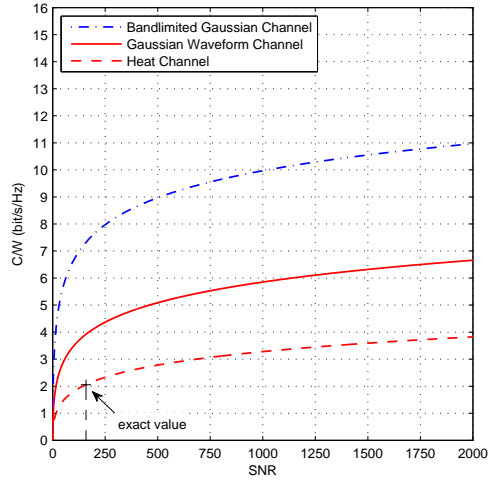}
  \caption{$C/W$ against SNR; \emph{exact value} refers to a sequence of random pulses as shown in 
  \cite[Fig. 7]{Poster} sent every time $2\pi\alpha$.}
  \label{Figure_2}
\end{figure}
\begin{figure}
  \centering
  \includegraphics[width=2.5in]{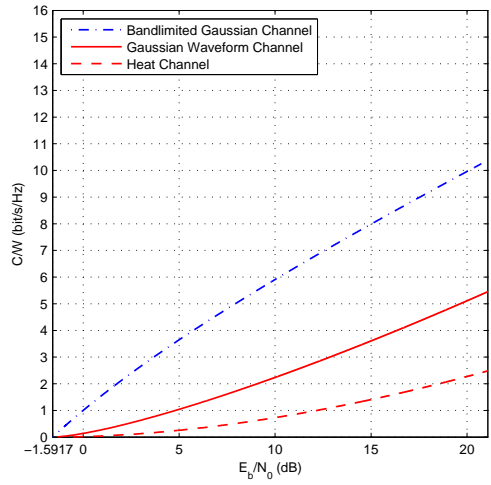}
  \caption{$C/W$ against $E_\mathrm{b}/N_0$.}
  \label{Figure_3}
\end{figure}
In Fig.~\ref{Figure_2}, $\bar{C}(\mathrm{SNR})/W$ is plotted against SNR; here, Eq.~(\ref{Eq2}) is 
used. In Fig.~\ref{Figure_3}, the curve for $\bar{C}(\mathrm{SNR})/W$ as a function of 
$E_\mathrm{b}/N_0$ is given parametrically by Eqs.~(\ref{Eq3a}),~(\ref{Eq3b}). In both figures, in case 
of the heat channel the label $C/W$ stands for $\bar{C}(\mathrm{SNR})/W$. Figs.~3 and 4 of the poster 
\cite{Poster} should be replaced by Figs.~2 and 3, resp., of the present publication. Moreover, the 
caption in \cite[Fig.~7]{Poster} is to be replaced by ``Example: $\alpha=50\,\mathrm{ps}$, 
$\beta= 200\,\mathrm{GHz}$, $\mathrm{SNR}=1000/(2\pi)=159.1\Rightarrow K=30,\,C=64.59$ bits per 
transmission (exact value)." The exact value, by the way, has been computed numerically.

\end{document}